\documentclass[prb,twocolumn,groupedaddress,a4paper,amsmath,amssymb,floatfix]{revtex4}
\usepackage{graphicx}
\usepackage{epstopdf}
\usepackage{color}

\def\openone{\leavevmode\hbox{\small1\kern-3.3pt\normalsize1}}

\newcommand{\Op}[1]{\boldsymbol{\mathsf{\hat{#1}}}}

\newcommand{\Fkt}[1]{\,\mathsf {#1}}

\ifx\Tr\renewcommand{\Tr}{\Fkt{Tr}}
\else\newcommand{\Tr}{\Fkt{Tr}}
\fi

\begin{document}

\title{Femtosecond two-photon photoassociation of hot magnesium atoms:\\
  A quantum dynamical study using thermal random phase wavefunctions}

\author{Saieswari Amaran}
\affiliation{Fritz Haber Research Centre and The
    Department of Physical Chemistry, Hebrew University, Jerusalem
    91904, Israel}

\author{Ronnie Kosloff}
\affiliation{Fritz Haber Research Centre and The
    Department of Physical Chemistry, Hebrew University, Jerusalem
    91904, Israel}

\author{Micha\l{} Tomza}
\affiliation{Department of Chemistry, University of
    Warsaw, Pasteura 1, 02-093 Warsaw, Poland}

\author{Wojciech Skomorowski\footnote[1]{Present address:
    Theoretische Physik, Universit\"at Kassel,
    Heinrich-Plett-Stra{\ss}e 40, 34132 Kassel, Germany
  }}
\affiliation{Department of Chemistry, University of
    Warsaw, Pasteura 1, 02-093 Warsaw, Poland}

\author{Filip Paw\l owski\footnote[2]{Also at:
    Physics Institute, Kazimierz Wielki University, Plac Weyssenhoffa 11,
    85-072 Bydgoszcz, Poland}}
\affiliation{Department of Chemistry, University of
    Warsaw, Pasteura 1, 02-093 Warsaw, Poland}

\author{Robert Moszynski}
\affiliation{Department of Chemistry, University of
    Warsaw, Pasteura 1, 02-093 Warsaw, Poland}

\author{Leonid Rybak}
\affiliation{The Shirlee Jacobs Femtosecond Laser Research Laboratory, Schulich Faculty of Chemistry, Technion-Israel Institute of Technology, Haifa 32000, Israel}

\author{Liat Levin}
\affiliation{The Shirlee Jacobs Femtosecond Laser Research Laboratory, Schulich Faculty of Chemistry, Technion-Israel Institute of Technology, Haifa 32000, Israel}

\author{Zohar Amitay}
\affiliation{The Shirlee Jacobs Femtosecond Laser Research Laboratory, Schulich Faculty of Chemistry, Technion-Israel Institute of Technology, Haifa 32000, Israel}

\author{J. Martin Berglund}
\affiliation{Theoretische Physik, Universit\"at Kassel,
  Heinrich-Plett-Stra{\ss}e 40, 34132 Kassel, Germany}

\author{Daniel M. Reich}
\affiliation{Theoretische Physik, Universit\"at Kassel,
  Heinrich-Plett-Stra{\ss}e 40, 34132 Kassel, Germany}

\author{Christiane P. Koch}
\affiliation{Theoretische Physik, Universit\"at Kassel,
  Heinrich-Plett-Stra{\ss}e 40, 34132 Kassel, Germany}
\email{christiane.koch@uni-kassel.de}

\begin{abstract}
  Two-photon photoassociation of hot magnesium atoms by femtosecond
  laser pulses, creating electronically excited magnesium dimer
  molecules, is studied from first principles,
  combining \textit{ab
    initio} quantum chemistry and molecular quantum dynamics.
  This theoretical framework allows for rationalizing the generation of
  molecular rovibrational coherence from thermally hot atoms
  [L. Rybak \textit{et al.}, Phys. Rev. Lett. {\bf 107}, 273001 (2011)].
  Random phase thermal wave functions are employed to model the
  thermal ensemble of hot colliding atoms. Comparing two
  different choices of basis functions, random
  phase wavefunctions built from eigenstates are
  found to have the fastest convergence
  for the photoassociation yield.
  The interaction of the colliding atoms with a
  femtosecond laser pulse is modeled non-perturbatively to account for 
  strong-field effects.
\end{abstract}

\maketitle

\section{Introduction}
\label{sec:intro}

Molecules can be assembled from atoms using laser light. This process is
termed photoassociation. With the advent of femtosecond lasers
and pulse shaping techniques, photoassociation became a natural
candidate for  coherent control of a binary
reaction.
Coherent control had been conceived as a method to determine the fate of
chemical reactions using laser fields.\cite{RonnieDancing89}
The basic idea is to employ interference of matter
waves to constructively enhance a desired outcome while destructively
suppressing all  undesired alternatives.\cite{RiceBook,ShapiroBook}
Control is exerted by shaping the laser pulses,
the simplest control knobs being time delays
and phase differences.\cite{DavidBook}
Over the last two decades,
the field of coherent control has developed significantly both
theoretically and
experimentally.\cite{GordonARPC97,BrixnerCPC03,DantusCR04,WollenhauptARPC05,SFB450book}
However, a critical examination of the
achievements reveals that successful
control has been demonstrated almost exclusively for
unimolecular processes such as ionization, dissociation and
fragmentation.
It is natural to ask why the reverse process of
controlling binary reactions\cite{MarvetCPL95,BackhausCP97,GrossJCP97,BackhausJPCA98,BackhausCPL99,ReginaFaraday99,BonnSci99,GeppertJCP03,NuernbergerPNAS10}
is so much more difficult.

The main difference between unimolecular processes and a binary
reaction lies in the initial state
-- a single or few well-defined bound quantum states vs an incoherent
continuum of scattering states.\cite{ZemanPRL04}
For a binary reaction, the nature of the
scattering continuum is mainly determined by
the temperature of the reactants.
As temperature decreases, higher partial waves
are frozen out. At the very low temperatures of ultracold gases,
the scattering energy of atom pairs is so low that
the rotational barrier cannot be passed, and the scattering becomes
purely $s$-wave.\cite{photoasso} In this regime, the reactants are
pre-correlated due to quantum threshold effects\cite{MyPRL09} and the
effect of scattering resonances is particularly
pronounced.\cite{PellegriniPRL08,AlyabyshevPRA10,GonzalezKochPRA12}
At a temperature of about 100$\,\mu$K, photoassociation with
femtosecond laser pulses has been demonstrated.\cite{SalzmannPRL08}
Coherent transient Rabi oscillations were observed as the prominent
feature in the pump-probe spectra. The transients are due to long tails  of the
pulses caused by a sharp spectral cut which is necessary to avoid
excitation into unbound states.\cite{SalzmannPRL08,MerliMyPRA09}
This pinpoints to the fact that the large spectral bandwidth of a
femtosecond pulse is unsuitable to one-photon photoassociation
at ultralow temperatures. In this regime, a narrow-band
transition needs to be driven in order to avoid atomic
excitation.\cite{MyPRA06a,MyFaraday09,TomzaKochPRA12}

The situation changes completely for high
temperatures where the scattering states can penetrate rotational
barriers due to the large translational kinetic energy. The association
process is then likely to happen at short internuclear distance close
to the inner turning point and for highly excited rotational
states. In this case, the large spectral bandwidth of femtosecond
laser pulses is ideally adapted to both the broad thermal width of the
ensemble of scattering states and the depth of the electronically
excited state potential in which molecules are formed. The
disadvantage of this setting is that the initial state is completely
incoherent, impeding control of the photoreaction.
Photoassociation with femtosecond laser
pulses was first demonstrated  under these
conditions, employing a one-photon transition in the
UV.\cite{MarvetCPL95} Subsequent to the photoassociation,
coherent rotational motion of the molecules was
observed.\cite{MarvetCPL95} We have recently demonstrated
generation of both rotational and vibrational coherences by two-photon
femtosecond photoassociation of hot
atoms.\cite{RybakPRL11,RybakFaraday}
This is a crucial step toward the coherent control of photoinduced
binary reactions since the fate of bond making and
breaking is determined by the vibrational motion.

Employing multi-photon transitions comes with several
advantages:  The class of molecules that can be
photoassociated by near-IR/visible femtosecond laser pulses is
significantly larger for multi-photon than  one-photon
excitation. Femtosecond laser technology is most advanced in the
near-IR spectral region.
Due to the different selection rules, different
electronic states become accessible for multi-photon transitions
compared to one-photon excitation. Control strategies differ for
multi-photon and one-photon excitation. In particular, large dynamic
Stark shifts and an extended manifold of quantum pathways that can be
interfered come into play for multi-photon excitation.\cite{YaronARPC09}
The theoretical description needs to account for these strong-field effects.

\begin{figure}[tb]
  \centering
  \includegraphics[width=8.7cm]{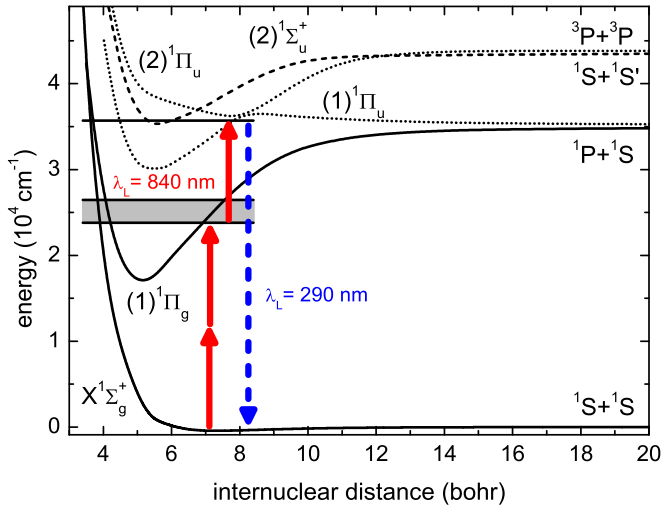}
  \caption{Potential energy curves of the electronic states involved
    in the two-photon photoassociation probed by a time-delayed
    pulse. The shaded region indicates the vibrational band populated
    after photoassociation.
  }
  \label{fig:scheme}
\end{figure}

We have constructed a comprehensive theoretical model from first principles
to describe the experiment in which magnesium atoms in a
heated cell are photoassociated by femtosecond laser
pulses.\cite{RybakPRL11,RybakFaraday}
It is summarized in Figure~\ref{fig:scheme}.
Magnesium in its electronic ground state is a closed shell atom.
Its ground electronic potential, X$^1\Sigma_g^+$, therefore displays
only a weak van der Waals attractive well. A
femtosecond pulse of 100$\,$fs transform-limited duration with a
central wavelength of $\sim 840\,$nm promotes
an electron to the $\pi$ orbital.
This two-photon transition is driven since
a wavelength of 840$\,$nm is far from any one-photon resonance
both for magnesium atoms and Mg$_2$ molecules, cf. Fig.~\ref{fig:scheme}.
Upon excitation, a strong chemical bond
is formed in the $(1)^1\Pi_g $ state with a binding energy of  $\sim
1.8\,$eV or, equivalently, $\sim 14500\,$cm$^{-1}$.
A time-delayed femtosecond pulse probes the excited Mg$_2$ molecule
by inducing a one-photon transition to a higher excited electronic state
($^1\Pi_u$). This state has a strong one-photon transition back to
the ground state. The corresponding experimental observable is the
intensity of the resulting UV fluorescence ($\sim 290\,$nm), measured
as a function of the pump-probe time delay. An oscillating signal is a
manifestation of coherent rovibrational dynamics in the $^1\Pi_g $
state.\cite{RybakPRL11,RybakFaraday}

The correct description of the thermal initial state is
crucial to capture the generation of coherence out of an incoherent
ensemble. The density operator, $\Op\rho_T$, describing the initial
state of hot atom pairs at temperature, $T$, is constructed by a
thermal average over suitable basis functions. Since no dissipative
processes occur on the  sub-picosecond timescale of the experiment,
the coherent time evolution of the density operator is efficiently
carried out by propagating the basis functions. Expectation values
are obtained by thermally averaging the corresponding operator over
the propagated basis functions. A numerically efficient description
of the initial thermal ensemble is  essential to facilitate the
time-dependent simulations. The present work on \textit{ab initio} simulation of ultrafast
hot photoassociation presents a detailed account of theoretical and numerical components
and their integration  into a comprehensive framework.

The paper is organized as follows. Section~\ref{sec:model}
presents the theoretical framework by introducing the Hamiltonian
describing the coherent interaction of an atom pair with strong
femtosecond laser pulses. The
relevant electronic states, their potential
energy curves, transition matrix elements and non-adiabatic
couplings, all obtained employing highly accurate state of the art
\textit{ab initio} methods,  are discussed. Section~\ref{sec:thermal}
derives an effective description of the thermal ensemble of
translationally and rotationally hot atom pairs in their electronic
ground state based on random phase thermal wave functions. We consider
three different choices of basis functions, two of them turn out to be
practical.
Convergence of the photoassociation probability
is studied in Section~\ref{sec:conv} for the different thermal
averaging procedures, and the role of shape resonances is discussed.
Section~\ref{sec:results} investigates the
generation of  coherence  in terms of the quantum purity and a
dynamical coherence measure. Finally, we conclude in
Section~\ref{sec:concl}.
Atomic units are used throughout our paper, unless specified
otherwise.

\section{\textit{ab initio} Model}
\label{sec:model}

The coherent (2+1) three-photon excitation of a pair of magnesium
atoms, that collide with rotational quantum number $J$,
by a strong femtosecond laser pulse is described by the
time-dependent Hamiltonian,
\begin{widetext}
  \begin{equation}
    \label{eq:Htot}
    \Op H^J(t) =
    \begin{pmatrix}
      \Op H^J_{X^1\Sigma_g^+}+\omega^S_{X^1\Sigma_g^+}(t,R)
      & \chi^\star(t,R) & 0 & 0
      & 0 & 0 \\
      \chi(t,R) & \Op
      H^J_{(1)^1\Pi_{g}}+\omega^S_{(1)^1\Pi_g}(t,R)
      & W_{1}(R) &
      W_{2}(R) & \mu_1(R) E^\star(t) & \mu_2(R) E^\star(t)\\
      0 & W_{1}(R) & \Op H^J_{(1)^3\Sigma^+_{g}} & W_{3}(R) & 0 &0 \\
      0 & W_{2}(R) & W_{3}(R) & \Op H^J_{(1)^3\Pi_{g}} & 0 & 0 \\
      0 & \mu_1(R) E(t) & 0 & 0 & \Op{H}_{11}^J+\omega^S_{11}(t,R)  &
      V_{12}(R)+\omega^S_{12}(t,R) \\
      0 & \mu_2(R) E(t) & 0 & 0 & V_{12}(R)+\omega^S_{12}(t,R)
      & \Op{H}_{22}^J+\omega^S_{22}(t,R) \\
    \end{pmatrix}\,.
  \end{equation}
\end{widetext}
Here $\Op H^J_{a}$
is the nuclear Hamiltonian of electronic state $a$,
\begin{equation}
  \label{eq:Ha}
  \Op H^J_{a} = \Op T + V_{a}(R) +\frac{J(J+1)}{2mR^2}\,,
\end{equation}
with $\Op{T} = \Op{P}^2 /2 m$ the
vibrational kinetic energy, $m$ the reduced mass and
$V_a(R)$ the potential energy curve of electronic state $a$.
$\mu_1$ and $\mu_2$ denote the (one-photon) transition dipole moments
between the
$(1)^1\Pi_g$ state and the first and second $^1\Pi_u$ states.
The Hamiltonian~\eqref{eq:Htot} neglects ro-vibrational couplings.
In a two-photon rotating-wave approximation,
the two-photon coupling between the $X^1\Sigma_g^+$ $(g)$ and $(1)^1\Pi_{g}$
$(e)$ states is denoted by $\chi(t, R)$,\cite{BoninBook}
\begin{equation}
  \label{eq:chi}
  \chi(t,R)=\frac{1}{4} E(t)^2 \sum_{i,j}
  \epsilon_i\epsilon_jM_{ij}^{e\gets g}(R),
\end{equation}
with $E(t)=S(t)e^{i\varphi(t)}$ the electric field envelope of the
laser pulse, $\epsilon_i$ the polarization component
($i=\perp,\parallel$),  and $M_{ij}^{e \gets g}$
the tensor elements of the two-photon electric transition
dipole moment between the ground ($g$) and excited ($e$) states,\cite{BoninBook}
\begin{equation}
\label{eq:M_ij}
M_{ij}^{e \gets g}(R)=-\sum_{n}\left[
  \frac{\langle e|\Op{\mu}_i|n\rangle
    \langle n|\Op{\mu}_j|g\rangle}
  {\omega_{ng}-\omega_L}+
  \frac{\langle e|\Op{\mu}_j|n\rangle
    \langle n|\Op{\mu}_i|g\rangle}{\omega_{ne}+\omega_L}
\right]\,.
\end{equation}
The summation is carried out over all electronic states $n$, 
except for the states which are explicitly accounted for in our model, 
cf. Eq.~\eqref{eq:Htot}. 
$\omega_{ng}$ and $\omega_{ne}$ are the transition frequencies 
between state $n$ and, respectively, state $g=X^1\Sigma_g$ and
$e=^1\Pi_g$.
Note that the two-photon transition moment, $M_{ij}^{e \gets g}(R)$,
depends on the central laser frequency, $\omega_L=hc/\lambda_L$. Here
we keep $\lambda_L=840\,$nm fixed. The strong laser field
driving the two-photon transitions may lead to non-negligible
dynamic Stark shifts $\omega^S_a(t,R)$,\cite{BoninBook}
\begin{equation}
\omega^S_a(t,R)=-\frac{1}{4}|E(t)|^2\sum_{i,j}
\epsilon_i\epsilon_j\alpha^a_{ij}(\omega_L,R)\,,
\end{equation}
where the tensor elements of the dynamic electric dipole
polarizability are given by\cite{BoninBook}
\begin{equation}
\label{eq:alpha}
\alpha^a_{ij}(R)=\sum_{n\ne a}\left[
  \frac{\langle a |\Op\mu_i|n\rangle
    \langle n|\Op\mu_j|a\rangle}
  {\omega_{na}-\omega_L}
  +\frac{\langle a|\Op\mu_j|n\rangle
    \langle n|\Op\mu_i|a\rangle}{\omega_{na}+\omega_L}\right]\,,
\end{equation}
where the sum runs over all electronic states
$n$, except those explicitly accounted for in our model,
cf. Eq.~\eqref{eq:Htot}, and 
$\omega_{na}$ is the transition frequency between states $n$ and $a$
($a=e,g$). 
We account only for the isotropic
part of the polarizability, neglecting anisotropic terms that occur
for open shell states with the projection of the electronic angular
momentum not equal to zero.\cite{SkomorowskiJCP11,TomzaMolPhys13}
This corresponds to
two-photon transitions with $\Delta J=0$, neglecting transitions with
$\Delta J = \pm 2$.
Similarly to the two-photon transition moment,  $M_{ij}^{e \gets g}(R)$,
the dynamic polarizability, $\alpha^a_{ij}(R)$,
depends on the central laser frequency, $\omega_L$.
Note that resonant transitions, both one-photon and two-photon
transitions, are treated in a non-perturbative way while all
non-resonant transitions are accounted for within second order
perturbation theory. 

The $(1)^1\Pi_g$ excited state that is accessed by the two-photon
transition is weakly coupled to the $(1)^3\Pi_g$, and
$(1)^3\Sigma_g$ states due to the spin-orbit interaction.
The spin-orbit matrix elements relevant for our work read
\begin{equation}
W_1(R) = \langle\Psi_{(1)^1\Pi_g}|H_{\rm SO}| \Psi_{(1)^3\Sigma_g}\rangle,
\label{SO1}
\end{equation}
\begin{equation}
W_2(R) = \langle\Psi_{(1)^1\Pi_g}|H_{\rm SO}| \Psi_{(1)^3\Pi_g}\rangle,
\label{SO2}
\end{equation}
\begin{equation}
W_3(R) = \langle\Psi_{(3)^1\Pi_g}|H_{\rm SO}| \Psi_{(1)^3\Sigma_g}\rangle,
\label{SO3}
\end{equation}
where $H_{\rm SO}$ is the spin-orbit coupling Hamiltonian in the Breit-Pauli
approximation including all one- and two-electron terms.
The effect of the spin-orbit coupling was actually observed in the fluorescence
signal, but it was so weak that we could neglect the triplet states in the
time-dependent calculations.
A one-photon transition connects the $(1)^1\Pi_g$ state to the
adiabatic $(1)^1\Pi_u$ and  $(2)^1\Pi_u$ states that are strongly coupled by the
radial nuclear momentum operator. In order
to include this non-adiabatic coupling, the diabatic representation
is employed, see e.g. Ref.~\onlinecite{BaerBook}.
$V_{11}^{\rm d}$ and $V_{22}^{\rm d}$ denote the
corresponding diagonal diabatic
potentials and $V_{12}(R)=V_{21}(R)$ the coupling term. Analogously,
$\omega^S_{ij}(t,R)$ ($i,j=1,2$) denote the Stark shifts in the
diabatic basis.
The angle of the rotation matrix transforming adiabatic into diabatic
representation is given by\cite{BaerBook}
\begin{equation}
\zeta(R)=\int_R^\infty \tau(R')dR'
\end{equation}
with the nonadiabatic radial coupling
\begin{equation}
\tau(R)=\left\langle \Psi_{(1)^1\Pi_u}\left|
\frac{\mathrm{d}}{\mathrm{d} R} \right|
\Psi_{(2)^1\Pi_u}\right\rangle \,.
\end{equation}
Consequently, the one-photon transition dipole moments $\mu_1(R)$,
$\mu_2(R)$ are calculated from the diabatic molecular wave functions,
obtained by rotating the adiabatic  $(1)^1\Pi_u$ and  $(2)^1\Pi_u$
wave functions.

\begin{table}[th]
\caption{Spectroscopic characteristics, i.e.,
  equilibrium bond lengths, $R_e$, and well depths, $D_e$,
  of our {\em ab initio} potentials. \label{tab:spec}}
\begin{ruledtabular}
\begin{tabular}{lrrr}
state & $R_e$ (bohr) & $D_e$ (cm$^{-1}$) & Dissociation \\
\hline
$X^1\Sigma_g^+$ & 7.33 & 430 & $(1)^1S+(1)^1S$  \\
\hline
$(1)^3\Sigma_g^+$ & 12.80  & 49 & $(1)^1S+(1)^3P$ \\
$(1)^3\Pi_g$  & 5.31  & 7963 & $(1)^1S+(1)^3P$  \\
$(1)^3\Sigma_u^+$  & 5.72  & 7459 & $(1)^1S+(1)^3P$  \\
$(1)^3\Pi_u$  & 8.60 & 110 & $(1)^1S+(1)^3P$ \\
\hline
(2)$^1\Sigma_g^+$    &  6.22 &   2221    & $(1)^1S+(1)^1P$ \\
(1)$^1\Pi_g$         &  5.10 &  18077    & $(1)^1S+(1)^1P$  \\
A$^1\Sigma_u^+$      &  5.75 &   9427    & $(1)^1S+(1)^1P$  \\
(1)$^1\Pi_u$         &  5.50 &   5395    & $(1)^1S+(1)^1P$  \\
\hline
(3)$^1\Sigma_g^+$    &  5.00 &   6203    & $(1)^1S+(2)^1S$ \\
(2)$^1\Sigma_u^+$    &  5.57 &   8262    & $(1)^1S+(2)^1S$ \\
\end{tabular}
\end{ruledtabular}
\end{table}

State-of-the-art {\em ab initio} techniques have been applied to compute the
potential energy curves of the magnesium dimer in the Born-Oppenheimer
approximation. All calculations employed the aug-cc-pVQZ basis set of
quadruple zeta quality as the atomic basis for Mg.
This basis set was augmented by the set of bond functions
consisting of $[3s3p2d2f1g1h]$ functions placed in the middle of the
Mg dimer bond.
All potential energy curves were obtained by a supermolecule method, and
the Boys and Bernardi scheme was used to correct for the basis-set
superposition error.\cite{Boys:71}

The ground $X^1\Sigma_g^+$ state potential was computed with the coupled cluster
method restricted to single, double, and noniterative triple excitations, CCSD(T).
For the excited $^1\Pi_g$ and $(1)^1\Pi_u$ states, linear response theory
(equation of motion approach) within the coupled-cluster singles and
doubles  framework, LRCCSD, was employed.
The potential energy curve of the excited $(2)^1\Pi_u$ state in the region of the
minimum of the potential was also obtained with the LRCCSD method. At
larger internuclear distances this potential energy curve was represented by the
multipole expansion with electrostatic and dispersion terms $C_n/R^n$ up to
and including $n=10$. The long-range coefficients $C_n$ were obtained
within the multireference configuration interaction method restricted to
single and double excitations, MRCI, with a large active space. The latter
procedure was necessary since
the $(2)^1\Pi_u$ state dissociates into Mg$(^3{\rm P})$+Mg$(^3{\rm P})$ atoms
and cannot be asymptotically described  by a single Slater determinant.
The CCSD(T) and CCSD calculations, including the response functions
calculations, were performed with the \textsc{dalton}
program,\cite{dalton} while the MRCI
calculations were carried out with the \textsc{molpro} suite of
codes.\cite{molpro}

The energy of the separated atoms was set equal to the experimental
value for each electronic state, although the atomic excitation
energies obtained from the LRCCSD calculations
were very accurate and for the lowest $^1$P state the deviation
from the experimental values was approximately 100$\,$cm$^{-1}$.
A high accuracy of the computed potential energy curves is confirmed by an
excellent agreement of the theoretical dissociation energy for the
ground $X^1\Sigma_g^+$ state ($D_0=$403.1cm$^{-1}$) with the experimental
value ($D_0=$404.1$\pm$0.5cm$^{-1}$).\cite{Balfour:70}
Moreover, the number of bound vibrational states for $J=0$ supported by
the electronic ground state agrees with the experimental number, $N_\nu=19$.
Spectroscopic parameters of the other experimentally observed state,
$A^1\Sigma_u^+$,
also agree with our values, for the well position within 0.07$\,$Bohr,
while the binding energy ($D_e$=9427cm$^{-1}$) is only 0.4\% higher than the
experimental value ($D_e$=9387cm$^{-1}$).\cite{Balfour:70}
The  root mean square deviation of the rovibrational levels
computed with the potential energy
curves from the CCSD(T) and LRCCSD calculations for the ground and $A$
states were 1.3$\,$cm$^{-1}$ and
30$\,$cm$^{-1}$, respectively, i.e., 0.3\% of the potential well depth.
Such a good agreement of the calculations with the available experimental data
strongly suggests that we can expect a comparable level of accuracy
for the other computed potential energy curves and molecular properties.

\begin{figure}[tbp]
  \centering
  \includegraphics[width=\linewidth]{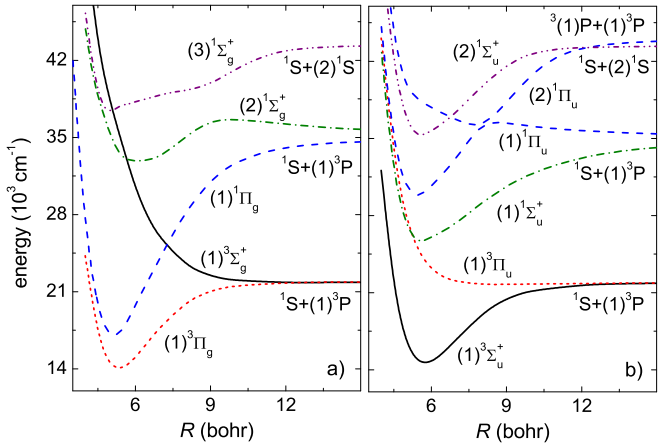}
  \caption{Potential energy curves of the gerade (left panel) and ungerade (right panel)
excited states of the magnesium dimer.}
  \label{fig:abinitio}
\end{figure}

The spectroscopic characteristics of the ground and excited electronic states
are gathered in Table~\ref{tab:spec}, while the corresponding potential energy
curves are reported in Fig.~\ref{fig:abinitio}. Inspection of Table~\ref{tab:spec}
shows that most of the excited electronic states of Mg$_2$ are strongly bound with
the dissociation energies ranging from 5400$\,$cm$^{-1}$ for the $(1)^1\Pi_u$ state
up to 18000$\,$cm$^{-1}$ for the $(1)^1\Pi_g$ state. Only
the $(1)^3\Sigma_g^+$ and $(1)^3\Pi_u$ states are very weakly bound with
binding energies of 49$\,$cm$^{-1}$ and 110$\,$cm$^{-1}$, respectively.
The agreement of our results with data reported by Czuchaj and
collaborators in 2001\cite{Czuchaj:01} is relatively good, given the fact that their
results were obtained with the internally contracted multireference
configuration singles and doubles method based on a CASSCF reference function.
Indeed, for the $(1)^1\Pi_g$, A$^1\Sigma_u^+$, $(1)^3\Sigma_g^+$,
$(1)^3\Pi_g$, $(1)^3\Pi_u$, and $(1)^3\Sigma_u^+$ states
the computed well depths agree within 600 cm$^{-1}$ or better,
i.e., within a few percent, while the equilibrium distances agree within
a few tenths of bohr at worst. Only for the $(1)^1\Pi_u$ state
we observe a very large difference in the binding energy, 3000 cm$^{-1}$. Such
a very strong binding in the $(1)^1\Pi_u$ state is very unlikely, since this state
would then show a strong interaction with the spectroscopically observed
A$^1\Sigma_u^+$ state. This interaction would, in turn, show up in the observed
${\rm A}^1\Sigma_u^+ \leftarrow {\rm X}^1\Sigma_g^+$ spectra as inhomogeneous
perturbations of lines. However, such perturbations have not been observed in the
recorded spectra,\cite{Balfour:70} suggesting our {\em ab initio} potential
for the $(1)^1\Pi_u$ state to be more accurate.

\begin{figure}[tbp]
  \centering
  \includegraphics[width=\linewidth]{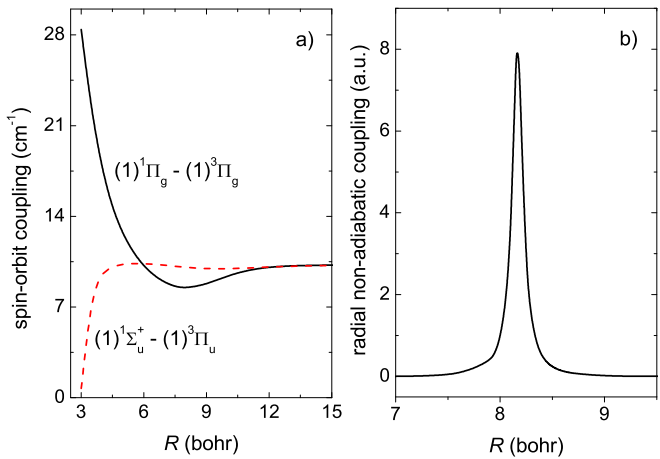}
  \caption{Spin-orbit coupling matrix elements between the $(1)^1\Pi_g$ and
the $(1)^3\Pi_g$ states, and $(1)^1\Sigma_u^+$ and $(1)^3\Pi_u$ states
(left panel) and radial nonadiabatic coupling between the$(1)^1\Pi_u$
and $(2)^1\Pi_u$ states of the Mg$_2$ molecule.}
  \label{fig:abinitio1}
\end{figure}
Potential energy curves for all electronic states computed in the present work
are reported in Fig. \ref{fig:abinitio}. They all show a smooth behavior with
well defined minima and in some cases maxima due to the first-order resonant
interactions. The potential energy curves of the A$^1\Sigma_u^+$ and $(1)^3\Pi_u$
states and of the $(1)^1\Pi_g$ and $(1)^3\Sigma_g^+$ cross each other. These
crossings should experimentally be observed as a perturbation due to
the very weak, but non-zero spin-orbit coupling between the singlet
and triplet states. Indeed, the
experimental data on the ${\rm A}^1\Sigma_u^+ \leftarrow {\rm X}^1\Sigma_g^+$
transitions,\cite{Balfour:70} and the UV fluorescence spectra from the $(1)^1\Pi_g$
state \cite{RybakPRL11,RybakFaraday} confirm weak perturbations due to the spin-orbit
coupling. The corresponding matrix elements are shown in Fig. \ref{fig:abinitio1}.
Except for small interatomic distances they show a weak $R$ dependence and smoothly
tend to the atomic value.
The $(2)^1\Sigma_g^+$ and $(3)^1\Sigma_g^+$ show an avoided crossing, but the gap
between the two curves is so large that most probably no homogeneous
perturbations
will be observed in the spectra. By contrast, the $(1)^1\Pi_u$ and $(2)^1\Pi_u$
states show a very pronounced avoided crossing with a gap of a few wavenumbers.
This suggests a strong interaction between these states through the radial
nonadiabatic coupling matrix element. The shape of this element is shown in
the right panel of Fig. \ref{fig:abinitio1}.  As expected, the nonadiabatic
coupling matrix element is a smooth Lorenzian-type function, which, in the
limit of an inifintely close avoided crossing, becomes a Dirac
$\delta$-function. The height and width of the curve depends on the
strength of the interaction, and the small width of the coupling in
Fig.~\ref{fig:abinitio1}b suggests a strong nonadiabatic coupling.
It is gratifying to observe that the maximum on the nonadiabatic
coupling matrix element agrees well with the location of the avoided
crossing, despite the fact that two very different methods were
employed in the {\em ab initio} calculations. As discussed above, the
potential energy curves were shown to be accurate, so we are confident
that also the nonadiabatic
coupling matrix element are essentially correct.

The electric transition dipole moments between states $i$ and $f$,
$\mu_j=\langle \Psi_i|\hat\mu_j|\Psi_f\rangle$,
where the electric dipole operator, $\hat\mu_j=r_j$, is given by the
$j$th component
of the position vector and $\Psi_{i/f}$ are the wave functions for the
initial and final states, respectively, were computed as the first residue
of the LRCCSD linear response function for the $X^1\Sigma_g^+$, $^1\Pi_g$,
and $^1\Pi_u(1)$ states. For transitions to the $(2)^1\Pi_u$ state,
the MRCI method was employed. The two-photon transition moment,
Eq.~\eqref{eq:M_ij},
can in practice be obtained as a residue of the cubic response
function.\cite{Olsen:85} For transitions between the $X^1\Sigma_g^+$
and $^1\Pi_g$ states, $M_{ij}^{f \gets 0}(\omega_L,R)$ was computed as
a residue of the coupled cluster cubic response function with electric
dipole operators and wave functions within the CCSD
framework.\cite{Hattig:98a,Hattig:98b}
The tensor elements of the electric dipole polarizability of the
ground $X^1\Sigma_g^+$ state were obtained as the coupled cluster
linear response function with electric dipole operators and wave
functions  within the CCSD framework.\cite{Christiansen:98}
The dynamic polarizabilities of the excited states were computed as double
residues of the coupled cluster cubic response function with electric
dipole operators  and wave functions within the CCSD
framework.\cite{Hattig:98c,Hattig:98d}
The nonadiabatic radial coupling matrix elements as well as
the spin-orbit coupling matrix elements have been evaluated with the
MRCI method.

\begin{figure}[tbp]
  \centering
  \includegraphics[width=\linewidth]{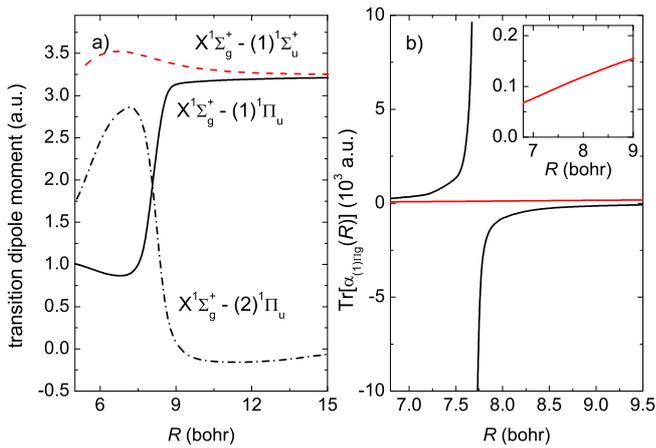}
  \caption{Electric transition dipole moments from the ground electronic state
    to the three lowest singlet states of ungerade symmetry (left
    panel), and dynamic Stark shift
    for the $(1)^1\Pi_g$ state as obtained from response calculations
    (black solid line)
    and after eliminating the contribution from $(2)^1\Sigma_u^+$ state
    (red dashed line) in
    the region important for the time-dependent calculations (right
    panel). The inset illustrates the smooth behavior of the dynamic
    Stark shift after elimination of the resonance due to the
    $(2)^1\Sigma_u^+$ state.
  }
  \label{fig:abinitio2}
\end{figure}
The electric transition dipole moments from the ground electronic state
to the three lowest singlet states of ungerade symmetry are reported
in Fig.~\ref{fig:abinitio2}. The transition moment to the A$^1\Sigma_u^+$
state is almost constant over a wide range of interatomic distances $R$,
and smoothly approaches the atomic value. The transition moments to the
$(1)^1\Pi_u$ and $(2)^1\Pi_u$ states show more pronounced variations.
In fact, the $R$ dependence of these two transition moments reflects
the avoided crossing of the corresponding potential energy curves
around $R=8.2$ bohr. Also reported in Fig.~\ref{fig:abinitio2} is the
trace of the dynamic Stark shift for the $(1)^1\Pi_g$ state as a
function of $R$. As expected from the definition \eqref{eq:alpha}, the
dynamic Stark shift shows resonances for transition energies close
to the laser frequency $\omega_L$. Since the adiabatic elimination
that leads to Eq.~\eqref{eq:alpha} assumes only non-resonant
transitions, the electronic states that cause
these resonances need to be included explicitly in the
non-perturbative Hamiltonian. This eliminates
their contribution to the dynamic Stark shift.
Fig.~\ref{fig:abinitio2}b illustrates the procedure,
showing the trace of the dynamic polarizability as a function of
$R$. The black solid line
shows a broad resonance around $R=7.7$ bohr.
This resonance corresponds to transitions from the
the $(1)^1\Pi_g$ to the $(2)^1\Sigma_u^+$  state. The latter state is
explicitly included in our Hamiltonian for the time-dependent
calculations and eliminated from Eq.~\eqref{eq:alpha}.
Once this is done, a smooth and almost constant behavior is obtained, as
illustrated in Fig.~\ref{fig:abinitio2}b. Of course, also the
contributions from all other electronic states that are explicitly
accounted for in the  Hamiltonian for the time-dependent
calculations are eliminated from Eq.~\eqref{eq:alpha}.

The Hamiltonian for the time-dependent calculation, neglecting the
weak spin-orbit coupling between the $(1)^1\Pi_g$, $(1)^3\Sigma_g^+$
and $(1)^3\Pi_g$ states and accounting for states with dipole
transitions that are near-resonant to the laser frequency, becomes
\begin{widetext}
  \begin{equation}
    \label{eq:Hpump}
    \Op H_{PA}^J(t) =
    \begin{pmatrix}
      \Op H^J_{X^1\Sigma_g^+}+\omega^S_{X^1\Sigma_g^+}(t,R)  & \chi^\star(t,R) & 0 & 0 & 0 \\
      \chi(t,R) & \Op H^J_{(1)^1\Pi_{g}} +\omega^S_{(1)^1\Pi_g}(t,R) &  \mu_1(R) E^\star(t) & \mu_2(R) E^\star(t) & \mu_3(R) E^\star(t) \\
      0 & \mu_1(R) E(t)  & \Op{H}_{11}^J+\omega^S_{11}(t,R)  &
      V_{12}(R)+\omega^S_{12}(t,R) & 0 \\
      0 & \mu_2(R) E(t) & V_{12}(R)+\omega^S_{12}(t,R)
      & \Op{H}_{22}^J+\omega^S_{22}(t,R) & 0 \\
      0 & \mu_3(R) E(t) & 0 & 0 & \Op H^J_{(2)^1\Sigma_u^+}
    \end{pmatrix}\,.
  \end{equation}
\end{widetext}
Note that Eq.~\eqref{eq:Hpump} makes use of the two-photon rotating wave
approximation, i.e., the electric field envelope, $E(t)$, in
Eq.~\eqref{eq:chi}, may be complex, $E(t)= E_0 S(t)
e^{i\varphi(t)}$, and a non-zero $\varphi(t)$ denotes the
relative phase with respect to the central laser frequency's phase.
The Hamiltonian~\eqref{eq:Hpump} is represented on an
equidistant grid for each partial wave $J$. Convergence with respect
to the grid size $R_{max}$ and number of grid points $N_R$ is
discussed below in Section~\ref{sec:conv}.

\section{Quantum dynamical description of a thermal ensemble}
\label{sec:thermal}

The initial state for photoassociation is given by the
ensemble of magnesium atoms in the heated cell which interact via the
$X^1\Sigma_g^+$ electronic ground state potential. Assuming
equilibrium, the initial state is represented
by the canonical density operator
for $N$ atoms held in a volume $V$ at temperature $T$. Due to the
moderate density in a heat pipe, the description can be restricted to
atom pairs.
The density operator for $N/2$ atom pairs is then obtained from
that for a single atom pair,
$\Op\rho_T(t=0)$, which is expanded into a suitable
complete orthonormal basis.\cite{MyJPhysB06}
Thermally averaged time-dependent expectation values of an
observable $\Op A$ are calculated according
to
\begin{equation}
  \label{eq:defA}
\langle \Op A\rangle_T(t) =\Tr[\Op A\Op \rho_T(t)]\,.
\end{equation}
The time evolution of $\Op\rho_T(t)$ is given by
$\Op\rho_T(t) = \Op U(t,0) \Op\rho_T(t=0) \Op U^+(t,0)$
starting from the initial state
\[
\Op\rho_T(t=0) = \frac{1}{Z}e^{- \Op H/k_BT}\,,
\]
where $\Op H$ is the Hamiltonian and $Z=\Tr[e^{-\Op H/k_BT}]$
the partition function.
For a thermal, i.e. incoherent, initial state, undergoing coherent
time evolution,  it is not necessary to solve the Liouville
von-Neumann equation for the density operator. Instead,
the dynamics can be captured by solving the Schr\"odinger equation
for each basis function.
Thermally averaged expectation values are calculated by properly
summing over the expectation values obtained from the propagated basis
states.\cite{MyJPhysB06}

Since many scattering states in a broad distribution of rotational
quantum numbers are thermally populated, the
approach of  propagating all thermally populated basis states
directly\cite{MyJPhysB06} becomes numerically
expensive. Alternatively,
an effective description of the thermal ensemble of scattering atoms
is obtained by averaging over realizations of random phases.
It makes use of thermal random
wave functions, $|\psi_{\alpha}^k\rangle$. Here, the index $k$ labels a set of
random phases and $\alpha$ the basis states.
Choosing an arbitrary complete orthonormal basis,
$\{|\alpha\rangle\}$, and given that
\[
\frac{1}{N}\sum_{k=1}^Ne^{i(\theta_\alpha^k-\theta_\beta^k)}=\delta_{\alpha\beta}
\]
for random phases  $\theta^k_\alpha$, $\theta^k_\beta$
and $N$ large, an expansion into random phase wave functions
yields a representation of unity,\cite{gelmanandronnie}
\begin{eqnarray}
  \label{eq:vONS}
  \openone = \frac{1}{N} \sum_{k=1}^N |\Psi^k\rangle\langle\Psi^k|
  &=& \frac{1}{N} \sum_{k=1}^N \sum_{\alpha\beta}
  e^{i(\theta_\alpha^k-\theta_\beta^k)}|\alpha\rangle\langle \beta| \nonumber\\
  &=& \frac{1}{N}\sum_{k=1}^N \sum_{\alpha\beta}
  |\Psi^k_\alpha\rangle\langle \Psi^k_\beta| \,,
\end{eqnarray}
where $|\Psi^k_\alpha\rangle=e^{i\theta_\alpha^k}|\alpha\rangle$ and
$|\Psi^k\rangle=\sum_\alpha e^{i\theta_\alpha^k}|\alpha\rangle$.

Here, we use a random phase expansion of unity for the 'vibrational'
degree of freedom, i.e., the radial part $R$ of the relative motion
$(R,\theta,\phi)$ of the diatom in its electronic ground state. No
electronic excitations are excited thermally.
A separation of rotational and vibrational dynamics and
subsequent expansion into partial waves is natural to make use of
spherical symmetry, $\Psi_{nJM}(R,\theta,\phi) =
\varphi_n(R|J) \otimes \langle\theta,\phi|J,M\rangle$, i.e.,
the vibrational motion is conditioned on $J$. This implies a complete
set of 'vibrational' basis functions (both true vibrational
eigenfunctions and scattering states), and subsequently, a different
set of random phases,  for each $J$.

In principle, one could apply a random phase expansion of unity also
in the rotational degree of freedom. This would be useful to study the
generation of rotational coherence. It requires a model that accounts
for rotational coherence, i.e., either a full rovibrational
Hamiltonian or, as a minimal approximation, a generalization of
Eq.~\eqref{eq:Hpump} comprising $\Op H^{J+2}_{(1)^1\Pi_g}$ and
$\Op H^{J-2}_{(1)^1\Pi_g}$ in addition to $\Op
H^J_{(1)^1\Pi_g}$. However, in the present work, we focus on the
generation of vibrational coherence which is crucial for bond
formation.

In the following, we
discuss three possible bases for the vibrational Hamiltonian,
from which random phase  wave functions
are generated. All three possibilities will lead, when averaged,
to the initial thermal ensemble corresponding to the experiment.
While formally equivalent, convergence of the
thermal averages with respect to the number of required
basis functions differs significantly for the three representations.

\subsection{Grid-based random phase approach}
\label{subsec:grid}

The simplest but, as it turns out, most inefficient approach uses, for
each partial wave $J$, the
coordinate basis of $\delta$-functions localized at each grid point $R$,
$\openone_J = \sum_R |R,J\rangle\langle R,J|$.\cite{gelmanandronnie}
A random phase wave function
is obtained by multiplying each basis state with a different random
phase, $\theta^k_{R,J}$,
\begin{equation}
  \label{eq:PsikR}
  |\Psi^k_{J}\rangle = \sum_{R=1}^{N_R} e^{i\theta_{R,J}^k}|R,J\rangle\,,
\end{equation}
with $k$ labeling one realization of $N_R$ random phases, $\{\theta^k_{R,J}\}$.
The resulting wave function, $\langle R|\Psi^k_{J}\rangle$,
has constant
amplitude and a different random phase at each $R$.
The initial density operator is obtained by propagating
each basis function $|\Psi^k_{J}\rangle$ under
$\Op H_g^J=\Op T + V_g(R) + \frac{J(J+1)}{2m R^2}$
in imaginary time, $\tau=\frac{i}{2}\beta$ with $\beta=1/k_BT$, using
the Chebychev propagator.\cite{RonnieImag86} This
yields the thermal random phase wave functions,
\begin{equation}
  \label{eq:PsikRT}
  |\Psi^k_{J}\rangle_T=e^{-\frac{\beta}{2}\Op H_g^J}|\Psi^k_{J}\rangle
\end{equation}
and thus the initial density operator,
\begin{widetext}
\begin{eqnarray}
\label{eq:rhoT}
\Op\rho_T(t=0) &=& \frac{1}{Z}e^{-\frac{\beta}{2} \Op
  H_g}e^{-\frac{\beta}{2} \Op H_g}
\frac{1}{N}\sum_{k=1}^N \sum_{R,R^\prime,J}(2J+1)
e^{i(\theta_{R,J}^k-\theta_{n^\prime,J^\prime}^k)}|R,J\rangle\langle R^\prime,J|
\nonumber \\
&=& \frac{1}{Z} \frac{1}{N}\sum_{k=1}^N\sum_{J=0}^{J_{max}} (2J+1)
e^{-\frac{\beta}{2} \Op H^J_g}|\Psi^k_{J}\rangle\langle\Psi^k_{J}|
e^{-\frac{\beta}{2} \Op H^J_g} =
\frac{1}{Z} \frac{1}{N}\sum_{k=1}^N\sum_{J=0}^{J_{max}} (2J+1)
|\Psi^k_{J}\rangle_T \,_T\langle\Psi^k_J| \,.
\end{eqnarray}
\end{widetext}
To calculate time-dependent expectation values,
$N(J_{max}+1)$ thermal random phase wave functions
$|\Psi^k_{J}\rangle_T$ are propagated in real time,
\begin{equation}
  \label{eq:propPsikTR}
  |\Psi^k_{J}(t)\rangle_T = \Op U(t,0)|\Psi^k_{J}\rangle_T  \,,
\end{equation}
with the
Hamiltonian~\eqref{eq:Hpump} as generator.
Thermally averaged time-dependent expectation values are obtained from
Eq.~\eqref{eq:defA},
using cyclic permutation under the trace,
\begin{widetext}
\begin{eqnarray}
  \label{eq:calcexp_R}
  \Tr \left[ \Op A \Op\rho_T(t) \right] &=&
  \frac{1}{N}\sum_{k=1}^N
  \sum_{R,J}(2J+1) 
  \langle \Psi^k_{R,J}|\Op A \Op U(t,0)
  \Op\rho_T(t=0)\Op U^+(t,0)|\Psi^k_{R,J}\rangle \nonumber \\ &=&
  \frac{1}{Z}\frac{1}{N}\sum_{k=1}^N
  \sum_{R,J}(2J+1) 
  \langle \Psi^k_{R,J}|
  e^{-\frac{\beta}{2}\Op H^J_g} \Op U^+(t,0) \Op A \Op U(t,0)
  e^{-\frac{\beta}{2}\Op H^J_g} |
  \Psi^k_{R,J}\rangle \nonumber \\ &=&
  \frac{1}{Z}\frac{1}{N}\sum_{k=1}^N
  \sum_{J=0}^{J_{max}}(2J+1)\; _T\langle \Psi^k_{J}(t)|\Op
  A|\Psi^k_{J}(t)\rangle_T \,.
\end{eqnarray}
\end{widetext}
Obtaining the $N(J_{max}+1)$ solutions of the Schr\"odinger equation,
$|\Psi^k_{J}(t)\rangle_T$, required by
Eq.~\eqref{eq:calcexp_R} requires typically significantly less
numerical effort than propagating
$(J_{max}+1)$ $N_R \times N_R$-dimensional density matrices,
neglecting the rovibrational coupling, or once the full
$N_R(J_{max}+1)\times N_R(J_{max}+1)$-dimensional density matrix.
Note that while $|\Psi^k_{J}\rangle_T$ has zero components
on all electronic states except the ground state,
$|\Psi^k_{J}(t>0)\rangle_T$ will be non-zero for all electronic
states due to the interaction with the field.
Relevant expectation values are the excited state population after the
pump pulse, possibly $J$-resolved. The corresponding operators are the
projectors onto the electronically excited state, $\Op
P_e=|e\rangle\langle e|$, and $\Op P^J_e$, i.e.,
\begin{equation}
  \label{eq:PeJ}
  \langle \Op P^J_e\rangle(t_f) = \frac{1}{Z}\frac{1}{N}
  \sum_{k=1}^N \sum_{J=0}^{J_{max}}
  (2J+1) |\langle e|\Psi^k_{J}(t_f)\rangle_T|^2\,.
\end{equation}

The convergence of this approach is  slow. The number of realizations
required to reach convergence was found to be much larger than the
number of grid points. The reason for the slow convergence is that
there is no preselection of those basis states that are most
relevant in the thermal ensemble. We will therefore not use this
method and have included it here only for the sake of completeness.

\subsection{Eigenfunction-based random phase approach}
\label{subsec:eigen}

A preselection of the relevant states becomes possible by
choosing the eigenbasis $|n,J\rangle$
of $\Op H_g^J$ and evaluating the trace only for
basis states with sufficiently large thermal weights,
$e^{-E_{n,J}/2k_BT} > \epsilon$ where $\epsilon$ is a
prespecified error.
The eigenfunction-based random phase wave functions are given by
\begin{equation}
  \label{eq:Psikeigen}
  |\Psi^k_{J}\rangle =  \sum_n e^{i\theta_{n,J}^k}|n,J\rangle\,,
\end{equation}
where $J$ denotes the partial wave and $n$ is the vibrational quantum
number in the bound part of the spectrum of $\Op H_g^J$ or,
respectively, the label of box-discretized continuum states.
It is straightforward to evaluate
the representation of the initial density operator in this basis,
\begin{widetext}
\begin{eqnarray} \label{eq:rho_ini_eigen}
\Op\rho_T(t=0) &=& \frac{1}{Z}e^{-\frac{\beta}{2} \Op
  H_g}e^{-\frac{\beta}{2} \Op H_g}
\frac{1}{N}\sum_{k=1}^N \nonumber
\sum_{n,n^\prime,J}(2J+1) 
e^{i(\theta_{n,J}^k-\theta_{n^\prime,J}^k)}|n,J\rangle\langle n^\prime,J|
\\  &=&
\frac{1}{Z}\frac{1}{N}\sum_{k=1}^N
\sum_{n,n^\prime,J}(2J+1) 
e^{-\frac{\beta}{2}E_{n,J}+i\theta_{n,J}^k}
\;  e^{ -\frac{\beta}{2}E_{n^\prime,J}-i\theta_{n^\prime,J}^k}
\; |n,J\rangle\langle n^\prime,J|\nonumber
\\ &=&
\frac{1}{Z}\frac{1}{N}\sum_{k=1}^N\sum_{J=0}^{J_{max}}
(2J+1) |\Psi^k_J\rangle_T\,_T\langle \Psi^k_J|
\end{eqnarray}
\end{widetext}
where $E_{n,J}$ denotes an eigenvalue of the partial wave ground state
Hamiltonian, $\Op H_g^J$, and the
$N(J_{max}+1)$ initial random phase wave functions are
given by
\begin{equation}
  \label{eq:PsikTeigen}
  |\Psi^k_{J}\rangle_T= \sum_n
  e^{-\frac{\beta}{2}E_{n,J}+i\theta_{n,J}^k}
  |n,J\rangle\,.
\end{equation}
Thermally averaged time-dependent expectation values are calculated
analogously to Eq.~\eqref{eq:calcexp_R},
where the time-dependent wave functions are obtained by
propagating the wave functions of Eq.~\eqref{eq:PsikTeigen} instead of
the initial states given by Eqs.~\eqref{eq:PsikR} and
\eqref{eq:PsikRT}.
For convenience, we use normalized random phase wavefunctions instead
of Eq.~\eqref{eq:PsikTeigen},
\begin{equation}
  \label{eq:PsikTeigennorm}
  |\tilde\Psi^k_{J}\rangle_T= \frac{1}{\sqrt{Z_J^{R_{max}}}}|\Psi^k_{J}\rangle_T\,,
\end{equation}
where $Z_J^{R_{max}}=\sum_n e^{-\beta E_{n,J}}$ and $R_{max}$
indicates the size of the box.
For the thermally averaged time-dependent expectation values, this
yields
\begin{equation}
  \label{eq:expP_J}
  \Tr \left[ \Op A \Op\rho_T(t) \right] = \frac{1}{N}
  \sum_{k=1}^N\sum_{J=0}^{J_{max}} P_J
  \langle\tilde\Psi^k_J(t)|\Op A|\tilde\Psi^k_J(t)\rangle
\end{equation}
with
\begin{equation}
  \label{eq:P_J}
  P_J = \frac{(2J+1)Z_J^{R_{max}}}{Z}
\end{equation}
the weight of the contribution of partial wave $J$.

The eigenfunction-based random phase approach requires diagonalization
of the $J_{max}+1$ partial wave ground state Hamiltonians, $\Op H_g^J$. Depending
on the time required for the propagation of each basis state, this
effort may very well be paid off by the much smaller number of basis
states that need to be propagated.

The partition function for the computational box of radius $R_{max}$ is
straightforwardly evaluated in the eigenbasis,
\[
Z_{box}^{R_{max}} =\sum_J (2J+1) Z_J =
\sum_{J=0}^{J_{max}}\sum_{n=0}^{n_{max}}(2J+1) e^{-E_{nJ}/k_BT}\,,
\]
where $n_{max}$, $J_{max}$ are chosen such that
$e^{-E_{n_{max}+1,J_{max}+1}/k_BT}<\epsilon$.
Since we are interested in high
temperatures, it is natural to compare the calculated partition
function $Z_{box}$ to its classical approximation,
\begin{equation}
  \label{eq:Zclass}
  Z_{cl} =
  \frac{4\pi^2}{h^3}\sqrt{\frac{2m\pi}{\beta}} \int dJ \,2J\, Z_{J,cl}^{R_{max}}
\end{equation}
with
\begin{equation}
  \label{eq:Z_J}
  Z^{R_{max}}_{J,cl} = R_{max} e^{-\beta\frac{J^2}{2mR_{max}^2}}
  - \sqrt{\frac{\pi\beta J^2}{2m}}\Fkt{erfc}\left(
    \sqrt{\frac{\beta J^2}{2mR_{max}^2}}\right)
\end{equation}
and $\Fkt{erfc}(x)=1-\Fkt{erf}(x)$.
The derivation of $Z_{cl}$ is given in Appendix~\ref{sec:app}.
For a temperature of 1000$\,$K, we find $Z_{box}$ and $Z_{cl}$ to agree
within less than 1\%.
Inserting the classical approximation of $Z^{R_{max}}_{box}$ and
  $Z_J^{R_{max}}$ into
Eq.~\eqref{eq:P_J}, we find $P_J$ to roughly correspond to the
normalized Boltzmann weight at the end of the grid.

\subsection{Freely propagated Gaussian random phase wave packets}
\label{subsec:Gauss}

The third approach avoids diagonalization of the partial wave ground
state Hamiltonians, $\Op H_g^J$, approximating them by the kinetic
energy, $\Op T$, only. This approximation is valid at high temperatures
where the kinetic energy of the scattering atoms is much larger than their
potential energy due to the inter-particle interaction.
It starts from a Gaussian wave paket positioned sufficiently far from
the interaction region. If the width of the wave packet is adjusted
thermally, projection onto energy resolved scattering wave functions
yields Boltzmann weights,
\begin{equation}
  \label{eq:GaussT}
  |\Psi_J^{R_{0}}\rangle_T = \frac{1}{(\sqrt{2 \pi}  \sigma_{R,T})^{1/2}}
  e^{-\frac{(R-R_0)^2}{2\sigma_{R,T}^2}}|R,J\rangle \,.
\end{equation}
The thermal width is given by
$\sigma_{R,T}=1/\sigma_{P,T}=1/\sqrt{2m/\beta}=1/\sqrt{2m k_B T}$,
and $R_{0} \gg R_V$ where $R_V$ is the interaction region.
The Fourier transformed wave packet,
\[
|\tilde\Psi_J^{R_{0}}\rangle_T= \frac{1}{(\sqrt{2 \pi}  \sigma_P)^{1/2}}
e^{-\frac{P^2}{2\sigma_{P,T}^2}+ i P R_{0} }|P,J\rangle\,,
\]
corresponds to eigenstates of the kinetic energy with Boltzmann
weights,\cite{JiriPRA01} i.e., we approximate
$e^{-\frac{\beta}{2}\Op H_g^J}|P,J\rangle \approx
e^{-\frac{\beta}{2}\frac{\Op P^2}{2m}}|P,J\rangle$.
Random phase wave functions can be generated from
Eq.~\eqref{eq:GaussT} by real-time propagation under $\Op H_g^J$ as
follows.
The time-evolved wave packet at time $\tau^k$ is written as
\begin{equation}
  \label{eq:randomphaseGauss}
  |\Psi_J^{R_{0}}(\tau^k)\rangle_T =
  \sum_{n >n_0} c_{nJ}
  e^{-\frac{\beta}{2} E_{n,J}-i E_{n,J} \tau^k +i \theta^0_{n,J}}|n, J \rangle\,,
\end{equation}
where expansion into the scattering states of the
finite computation box, i.e., the states $|n,J\rangle$ with positive
energy ($n>n_0$), has been used.
$\theta^0_{n,J}$ is an initial phase due to $R_0$.
Comparing to Eq.~\eqref{eq:PsikTeigen}, the random phases are given
by $\theta^{k}_{n,J}=-E_{n,J} \tau^k$.
For sufficiently large times,
$\tau^k \gg  \beta/2 $ and $ v \tau^k \gg R_0  $ which, with
$v=p/m=\sqrt{2E/m}=\sqrt{\beta/m}$  yields $\tau^k\gg R_0\sqrt{m/\beta}$,
the wave function will spread significantly and fill the interaction
region which is a prerequisite to correctly represent the thermal
density. A different set of phases is obtained by
propagating the Gaussian wave packet under $\Op H_g^J$ for a time
$\tau^{k'}$. For Mg$_2$ and $T=1000\,$K, the two
limits translate into $\tau^k\gg 4\,$fs and
$\tau^k\gg 300\,$ps for $R_0=35\,$a$_0$. For large grids, these
numbers grow correspondingly. Moreover, to reproduce the Boltzmann
ensemble not only qualitatively, but also quantitatively, the smallest
frequency difference between scatterings states in the computation box
needs to be resolved. This translates into even longer propagation
times.


Practically, a coordinate
grid based wave packet, Eq.~\eqref{eq:GaussT}, is propagated under
$\Op H_g^J$ by a Chebychev propagator  where each realization
corresponds to a different time $\tau^k$.
The density operator is constructed by
averaging over the  times, $\tau^k$, chosen randomly. Alternatively,
if the eigenvalues $E_{n,J}$ and eigenstates $|n,J\rangle$ are known,
the thermal random phase wavefunctions can be obtained simply by
projection on the $|n,J\rangle$, choice of a random phase and
reassembly of the Gaussian wavepacket from the random-phase
projections.
This avoids the very long propagation times required to faithfully
represent the Boltzmann ensemble. The method of choice,
diagonalization  of the ground state Hamiltonians, $\Op H_g^J$, and
subsequent projection, or propagation under $\Op H_g^J$
depends on the dimensionality of the problem due to the different
scaling of diagonalization and propagation. For a diatomic, the
diagonalization approach was found to be more efficient. For larger
systems, the propagation approach is expected to take over.

The convergence of random phase wave functions built on thermal
Gaussians with respect to the photoassocation yield
is comparatively fast, only a few realizations are
sufficient. The drawback of the procedure is that only the free part
of the initial wave functions is represented, leaving out the
interaction energy of the true scattering states as well as initial
population in bound or quasi-bound states.
Thermal expectation values are obtained according to
Eq.~\eqref{eq:expP_J}
where $|\tilde\Psi_J^k\rangle_T=e^{-i\Op H_g^J\tau^k}|\Psi^{R_0}_J\rangle_T$
is a (normalized) Gaussian random packet, freely propagated for random
times $\tau^k$. The probability $P_J$ for partial wave $J$,
Eq.~\eqref{eq:P_J}, needs to account
for the fact that the Gaussian is initially positioned at $R_0$, not
the edge of the grid. Therefore, using the classical
approximation,
$R_{max}$ in Eq.~\eqref{eq:P_J} needs to be replaced by $R_0$.

Another variant  utilizes $\delta$-functions in
momentum space,
$\openone_J = \frac{1}{N}\sum_{k=1}^N\sum_{P,P'} |P,J\rangle\langle P',J|$,
and add random phases, $\theta^k_{P,J}$, to the momentum components
directly,
\[
|\Psi^k_{P,J}\rangle = e^{i\theta^k_{P,J}}|P,J\rangle \,.
\]
The random phases, $\theta^k_{P,J}$, translate into
positions of the Gaussian, $R^k_{0}$. This procedure reconstructs the
correct density in the regions of flat potential
but fails in the interaction region and is therefore not employed here.

\subsection{Calculating the quantum mechanical purity of a thermal
  ensemble}
\label{subsec:rhosq}

The laser pulse excites a small fraction of the incoherent
ensemble of ground state atom pairs to the $(1)^1\Pi_g$ state and
further to the first and second $^1\Pi_u$ state.
This action corresponds to distillation and leads to higher
purity and coherence of the photoassociated
molecules.\cite{RybakPRL11}
In order to study the purity of the subensemble of diatoms in
the excited electronic state,
\begin{equation}
  \label{eq:purity}
  \mathcal{P}_e(t)=\Tr [\Op \rho^2_{T,e}(t)]\,,
\end{equation}
the normalized density
operator of electronic state $|e\rangle$, is formally constructed,
\begin{widetext}
\begin{eqnarray}\label{eq:rhoeT}
\Op \rho_{T,e}(t) &=&
\frac{1}{\langle \Op P_e\rangle(t)}\frac{1}{N} \sum_{k=1}^N
\sum_{J=0}^{J_{max}} P_J\, 
\Op P_e |\tilde\Psi^k_{J}(t)\rangle_T\, _T\langle
\tilde\Psi^k_{J}(t)| \Op P_e\,.
\end{eqnarray}
In the grid representation using $N_R$ grid points,
$\Op \rho_{T,e}(t)$ becomes a matrix of size $N_R\times N_R$,
\begin{eqnarray*}
\rho_{T,e}(R,R^\prime;t) &=& \frac{1}{\langle \Op P_e\rangle(t)}
\frac{1}{N} \sum_{k=1}^N
\sum_{J=0}^{J_{max}} P_J\, 
\tilde\Psi^{k,e}_{T,J}(R,t) \;\tilde\Psi^{{k,e}^\star}_{T,J}(R^\prime,t)\,,
\end{eqnarray*}
\end{widetext}
where $\tilde\Psi^{k,e}_{T,J}(R,t)=\langle R,e|\tilde\Psi^{k}_{J}(t)\rangle_T$
is the excited state projection of the ($k,J$)th propagated thermal random
phase wave function, $\langle R|\tilde\Psi^k_J\rangle_T$.
Since we expect to populate only a limited number of $(1)^1\Pi_g$ state
eigenfunctions, say $N_m$, it is computationally advantageous to transform the
excited state component of the propagated thermal wave functions into
the rovibrational eigenbasis, $|\varphi^e_{mJ}\rangle$, of the
$(1)^1\Pi_g$ state,
\[
\tilde\Psi^{k,e}_{T,J}(R,t) = \sum_{m=0}^{N_m-1} c^{k,T}_{mJ}(t) \varphi^e_{mJ}(R)\,,
\]
with
\[
c^{k,T}_{mJ}(t) = \int\tilde\Psi^{k,e}_{T,J}(R,t)\varphi^{e*}_{mJ}(R)\, dR \,.
\]
The resulting density matrix,
\begin{widetext}
\begin{eqnarray*}
  \rho_{T,e}^{m,m^\prime}(t) &=& \frac{1}{\langle \Op P_e\rangle(t)}
  \frac{1}{N} \sum_{k=1}^N
  \sum_{J=0}^{J_{max}}P_J\, 
  c^{k,T}_{mJ}(t) c^{{k,T}^\star}_{m^\prime J}(t)\,,  \quad
  m,m^\prime=0,\ldots,N_m-1\,,
\end{eqnarray*}
\end{widetext}
is only of size $N_m\times N_m$ and can
more efficiently be squared to obtain the purity. Moreover, this representation lends
itself naturally to the evaluation of the dynamical coherence measure.
In the eigenbasis, we can decompose the density operator into
its static (diagonal) and dynamic (off-diagonal) part,
$\Op\rho = \Op\rho_{stat}+\Op\rho_{dyn}$.
Such a decomposition has been motivated in the study of dissipative
processes, in particular by the fact that pure dephasing does
not alter the static part.\cite{BaninJCP94,BaninCP94}
The dynamical coherence measure,
\begin{equation}
  \label{eq:coherence}
  \mathcal{C}_e(t) = \Tr[\Op\rho^2_{T,e,dyn}(t)]\,,
\end{equation}
captures
the part of the purity that arises from the dynamical part of the
density operator.\cite{BaninJCP94,BaninCP94}

The purity of the excited state subensemble after the pump pulse,
$\mathcal{P}_e(t_f)$, shall be compared to the initial purity of the
whole ensemble (in the electronic ground state),
\[
\mathcal{P}_g(t=0) = \Tr [\Op \rho^2_{T}(t=0)]\,.
\]
To this end, but also to determine the photoassociation probability,
cf. Eq.~\eqref{eq:PeJ}, the partition function $Z$ needs to be
determined explicitly. We need to take into account that our
computation box represents only a small part of the experimental
volume. The total partition function is therefore given by
$Z = Z_{box} \frac{V}{V_{box}}$,
where $V_{box}=\frac{4}{3}\pi R_{max}^3=4.97\times 10^{-18}\,$cm$^{3}$
for $R_{max}=200\,$a$_0$ and $V$ the experimental
volume. Alternatively, the probability of a single atom in our
computation box is $p_{box}=\rho V_{box}$ with $\rho$ the experimental
density, $\rho=4.8\times 10^{16}$ atoms/cm$^3$.
The probability of finding two atoms in the box is then simply
$p^2_{box} = 5.7\times 10^{-2}$. Using Eq.~\eqref{eq:expP_J} with $\Op
A=\openone$, the purity of the initial state is obtained as
\begin{equation}
  \label{eq:purityGS}
  \mathcal P_g(t=0) = p_{box}^2 \sum_{J=0}^{J_{max}} P_J^2 \,,
\end{equation}
taking $Z=Z_{box}$ in Eq.~\eqref{eq:P_J} when evaluating $P_J$.

\section{Convergence of the thermal averaging procedures:
  Photoassociation probability}
\label{sec:conv}

The interaction of the atom pair with the laser field
is simulated by solving $N(J_{max}+1)$ time-dependent Schr\"odinger
equations,
\begin{equation}
  \label{eq:tdse}
  i\hbar\frac{\partial |\psi_{J}^k(t)\rangle_T}{\partial t} =
  \Op H^J_{PA}(t) |\psi_{J}^k(t)\rangle_T\,,
\end{equation}
for $k=1,\ldots,N$ and $J=0,\ldots,J_{max}$,
with a Chebychev propagator\cite{chebyprop}
and thermally averaging the solutions according to
Eq.~\eqref{eq:calcexp_R}. To reduce the computational effort, we
evaluate all sums over $J$ in steps of five.

\begin{figure}[tb]
\begin{center}
\includegraphics[width=0.95\linewidth]{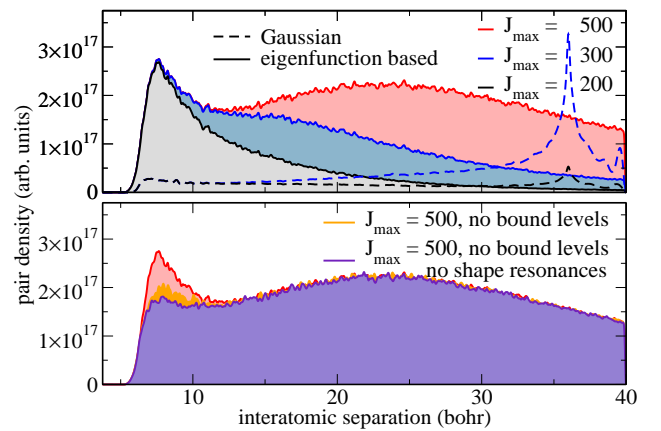}
\end{center}
\caption{Initial thermal density $\rho_{T,g}(R)/R^2$ of ground state
  atom pairs (calculated using Eq.~\eqref{eq:rho_ini_eigen} with 200
  realizations for each 
  $J$ and excluding bound states and shape resonances from the
  sum over $k$). 
}
\label{fig:rho_ini}
\end{figure}
We first study the initial thermal density of atom pairs,
cf. Eq.~\eqref{eq:rho_ini_eigen},  that is excited by the laser
pulse. It sets a limit to the excitation yield since thermalization
occurs over timescales larger than that of the experiment.
The initial thermal density of atom pairs is shown as a function of
interatomic distance in Fig.~\ref{fig:rho_ini} for random phase wave
functions built from eigenfunctions, cf. Eq.~\eqref{eq:PsikTeigennorm}
and built from Gaussians, cf. Eq.~\eqref{eq:randomphaseGauss}.
For photoassociation, distances smaller than $\sim 12\,a_0$
are relevant. The thermal density is converged in this region
by including rotational quantum numbers up to $J=300$. The
contribution of higher partial waves only ensures a constant density
at large interatomic distances.
The long-distance part naturally
converges very slowly but this is irrelevant for the dynamical
calculations. The peak at short interatomic
separations is due to bound levels, shape resonances and the classical
turning point of the scattering states at the repulsive barrier of the
potential: The difference between the red and orange curves in
Fig.~\ref{fig:rho_ini} indicates the contribution of bound levels, the
difference between the orange and the purple curve that of shape
resonances.

The Gaussian method requires much larger grids than the eigenfunction
based method to converge the initial thermal density
since it is based on the assumption that the effective
potential is zero at the position of the Gaussians. However, for large
values of $J$, the rotational barrier is non-zero even at
comparatively large internuclear separations. This leads to a spurious
trapping of probability amplitude, cf. the dashed curves in
Fig.~\ref{fig:rho_ini}. At short internuclear separations, the pair
density calculated from thermal Gaussians in the upper panel of
Fig.~\ref{fig:rho_ini} shows the same behavior as the purple curve in
the lower panel of Fig.~\ref{fig:rho_ini} (up to scaling which is due
to the accumulation of amplitude at large internuclear
separations). This indicates that random phase wave functions
built from thermal Gaussians do not capture bound states and shape
resonances.

\begin{figure}[tb]
\begin{center}
\includegraphics[width=0.9\linewidth]{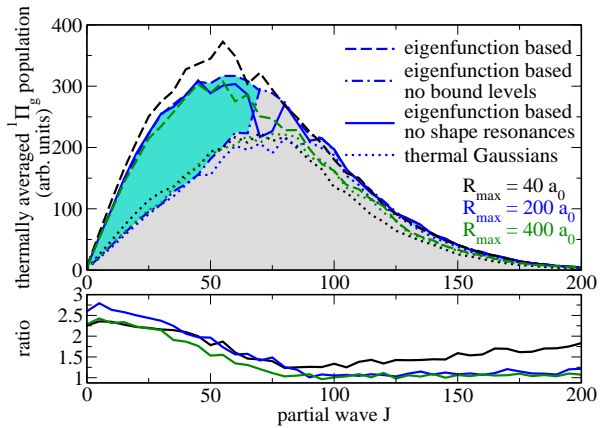}
\end{center}
\caption{Top panel:
  Thermally-averaged population of 
  the $^1\Pi_g$ state, calculated as
  $Z \langle \Op P_{e}\rangle$, vs initial partial wave $J$,
  averaged over 200 realizations of random phase wave functions, for
  three different grid sizes $R_{max}$ and a transform-limited pulse
  with $I_0=5\times 10^{12}\,$W/cm$^2$ and $\tau=100\,$fs (excluding
  the bound states and shape resonances from the 
  sum over $k$ in Eq.~\eqref{eq:expP_J}).
  The light-blue shaded part indicates the contribution of
  scattering states, i.e., photoassociation, for $R_{max}=200\,$a$_0$.
  The contributions of the shape resonances is given by the difference
  between the blue solid and dashed curves. Propagating random phase
  wave functions calculated from thermal Gaussians,
  cf. Eq.~\eqref{eq:randomphaseGauss},
  does not capture excitation of bound levels and shape resonances
  (dotted curves).
  Bottom panel: Ratio of the excitation yields calculated from
  random phase wavefunctions based on eigenfunctions and thermal
  Gaussians. For large $J$ and sufficiently large grid size, the two
  methods coincide as expected.
}
\label{fig:pumpfree}
\end{figure}
These two features of the Gaussian random phase wave functions show
also up in the population transferred from the initial incoherent
ensemble to the $^1\Pi_{g}$ state, shown in
Fig.~\ref{fig:pumpfree}.
The thermal averaging procedure has been repeated for increasing
initial rotational quantum number, $J$. That
is, for each rotational barrier,  eigenfunction-based and
Gaussian random phase wave functions are propagated in real time
with the full, time-dependent Hamiltonian, $\Op H_{PA}^J$.
Expectation values, such as the population of the
$^1\Pi_{g}$ state after the pump pulse is over, are calculated
for each random phase realization, $k$, and averaged over, including
the rotational
degeneracy factor $J+1$, cf. Eq.~\eqref{eq:calcexp_R}.
For large grids ($R_{max}=200\,$a$_0$, $R_{max}=400\,$a$_0$)
and large $J$, random phase wave functions built from
eigenfunctions and built from thermal Gaussians yield the same
results. Due to the trapping of probability amplitude at large
internuclear separations, for small grids ($R_{max}=40\,$a$_0$)
and large $J$, the Gaussian method underestimates
the excitation yield.
Since our random phase wave functions are normalized in the
computation box, this results in an initial thermal density which is too
small at the internuclear separations, $R\sim 7\,$a$_0\ldots 9\,$a$_0$,  that
are relevant for the laser excitation. This is illustrated by the
black curve in the lower panel of Fig.~\ref{fig:pumpfree} which
deviates from the blue and green curves even for large $J$.
For $J\le 75$, the potential supports bound levels which are
not captured by the Gaussian random phase wave functions.
Once the bound levels and shape resonances are removed from the
eigenfunction based approach (solid blue curve), the
eigenfunction-based approach roughly agrees with the Gaussian
approach (blue dotten curve). This comparison allows for estimating the
contribution of the bound levels. For $J\ge 75$ the ground state
potential does not support any bound levels due to the high
centrifugal barrier. The total contribution of the bound part of the
spectrum to the excitation of $^1\Pi_g$ population amounts to about
20\%.
The differences between $J=75$ to $J=95$ are attributed to insufficient sampling of the
free propagation method.

Qualitatively, however, the two approaches yield  the same result with
a steep rise at low $J$-values, a peak at intermediate $J$ and an
exponential tail for $J\ge 100$. The peak is shifted toward larger $J$
for the Gaussian method since it cannot capture the excitation of
bound levels and shape resonances.
Each random phase approach represents a statistical
sampling of the photoassociation yield.
The deviation of an expectation value
from its mean scales as $1/\sqrt{N}$ where $N$ is the number
of realizations. This was checked for  $J=55$ and $J=100$ the
pre-factor $\sigma/mean=\bar s/\sqrt{N}$ is estimated as
$\bar s \sim 0.37$ for the free propagation method and
$\bar s \sim 0.17$ for the eigenvalue method (note that $s \sim 0.30$
for the grid based method).
This makes the eigenvalue method converge fastest.

\begin{figure}[tb]
\begin{center}
\includegraphics[width=0.95\linewidth]{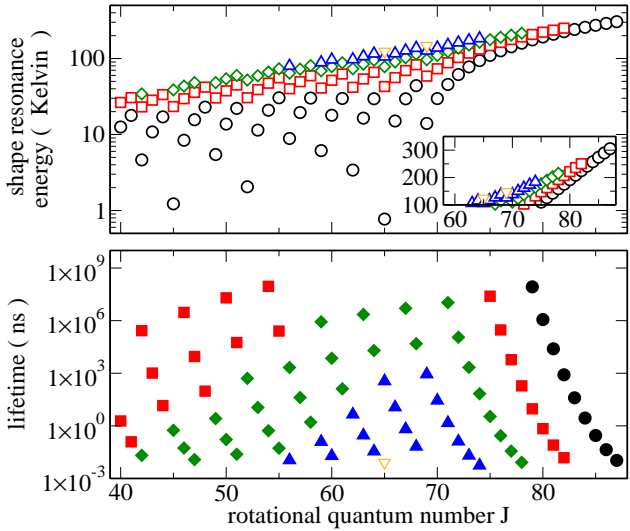}
\end{center}
\caption{Top panel: Position of shape resonances vs initial partial
  wave $J$. Bottom panel: Lifetimes in
  nanoseconds of the high-lying shape resonances vs partial wave
  $J$.
}
\label{fig:shaperes}
\end{figure}
The shape resonances are analyzed in
Fig.~\ref{fig:shaperes} which displays their position in energy (top
panel) and the lifetimes of those shape resonances that are
sufficiently short lived, i.e., sufficiently broad, to contribute to
the photoassociation process (bottom panel).
The shape resonances were calculated using a complex absorbing
potential.\cite{NimrodPhysRep98}
Shape resonances are
found for $J=40,\ldots,87$. The positions of the short-lived
resonances lie between 15$\,$K and 300$\,$K. With a sample
temperature of 1000$\,$K i.e., at least the higher lying of these
resonances are thermally populated. It is therefore not surprising
that a contribution of the shape resonances is observed
for $J=45\ldots,85$, cf. the difference between the solid and dashed
curves in Fig.~\ref{fig:pumpfree}. The contribution is easily
rationalized by the
shape resonances representing quasi-bound states that are ideally
suited for photoassociation.\cite{GonzalezKochPRA12}
They give structure to the
continuum of scattering states which otherwise is completely flat at
high temperature. This can be
utilized for generation of coherence and control.

In conclusion, the random phase wave functions built from thermal
Gaussians can be used if a rough estimate of the
photoassociation yield is desired.  When further refinement is required
the eigenvalue approach converges faster by a factor of 2.
Since the Gaussian approach excludes the bound part of
the spectrum and the resonances, it comes with error bars of about 20\%.
If more accurate results are desired, the eigenfunction
based random phase approach is the method of choice. The eigenfunction
based method is also best suited to capture the contribution of bound
states and shape resonances to the photoassociation yield.

\section{Interplay of coherence and photoassociation yield}
\label{sec:results}

\begin{figure}[tb]
\begin{center}
  \includegraphics[width=0.95\linewidth]{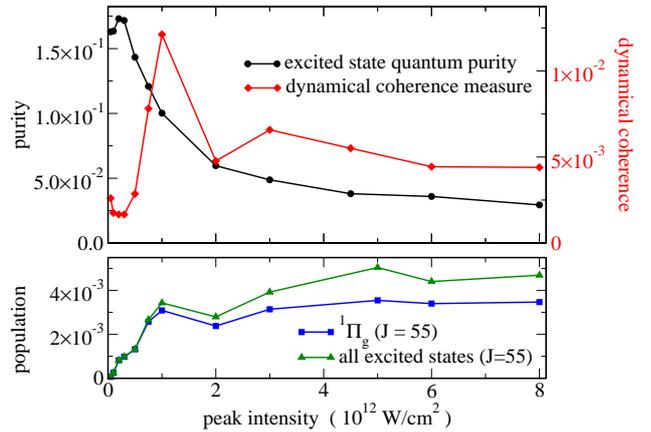}
\end{center}
\caption{Top panel:
  Excited state purity, $\mathcal{P}^{box}_e(t_{final})$, and coherence measure,
  $\mathcal{C}_e(t_{final})$, cf. Eqs.~\eqref{eq:purity} and
  \eqref{eq:coherence}, vs pulse intensity for the subensemble of
  photoassociated molecules. Bottom panel:
  Photoassociation yield for $J=55$ as a function of peak pulse
  intensity. All calculations employ transform-limited pulses of
  100$\,$fs full-width at half maximum.
}
\label{fig:intensity_j}
\end{figure}
The degree of distillation of coherence out of an incoherent initial
ensemble can be rationalized and quantified by considering the
enhancement of quantum purity  ${\cal P}=\Tr[\Op\rho^2]$ and
coherence. The experimental signature is the periodic modulation of
the pump probe signal.\cite{RybakPRL11,RybakFaraday}
The purity for an incoherent ensemble is inversely proportional to the
number of occupied quantum states of a pair of atoms.
This number is completely determined by the temperature
$T=1000\,$K, and density,  $\rho=4.8\cdot10^{16}\,$atoms/cm$^3$, in
the experiment. The bandwidth of our pulse plus Stark shift largely
exceeds the thermal width. This implies that all atom pairs within the
Franck-Condon window are excited on equal ground, irrespective of
their collision energy. 
We partition the total volume into identical smaller volumes
containing exactly one pair of atoms such that
the smaller volume corresponds to our computation box.\cite{MyJPhysB06}
The initial purity of the atom pair in our computation box of volume
$V_{box}$ is given by the ground state purity,
${\cal P}_g^{box}$, multiplied by the
probability for two atoms to
occupy this box, ${\cal P}^g=p^2 {\cal P}_g^{box}$,
cf. Section~\ref{subsec:rhosq}.
The initial ground state purity is bounded from below by the purity of a
maximally mixed state represented in our box. For $N=200$ random phase
realizations and $J_{max}=200/5$ we estimate the lower bound to be
$1\times 10^{-4}$. Evaluating ${\cal P}_g^{box}$ using the classical
approximation for $P_J$ in Eq.~\eqref{eq:purityGS}
we obtain  ${\cal P}_g^{box}=3.3\times 10^{-4}$ for
$R_{max}=200\,$a$_0$ and $T=1000\,$K
and thus
 ${\cal P}_g=1.9 \times 10^{-5}$ for the initial
purity.

For the ensemble of molecules in the electronically excited $^1\Pi_g$ state,
the density operator is given by Eq.~\eqref{eq:rhoeT} and the purity
within the computational box by Eq.~\eqref{eq:purity}. The actual
excited state purity is obtained by multiplying Eq.~\eqref{eq:purity}
by the probability for finding two atoms to be in the computational
box, $p^2$.
Note that the excited state density operator is normalized with
respect to the excitation yield, $\langle \Op P_e\rangle$.
We obtain a purity ${\cal P}_e^{box}\approx 5 \cdot 10^{-2}$
for the molecular sub-ensemble in the $^1\Pi_g$
excited state for the experimental pulse parameters. We thus observe a
significant increase in the quantum purity, $\Tr[\Op\rho^2]$,
induced by the femtosecond laser pulse.
The underlying physical mechanism can be viewed as "Franck-Condon
filtering":
for a given initial $J$ value there is only a limited range of
collision energies that allow the colliding pair to reach
the Franck-Condon window for PA located at short internuclear
distances\cite{BackhausCP97}.

In order to obtain a quantitative estimate of the degree of
distillation achieved by the femtosecond photoassociation process,
we have calculated the purity of the ensemble of photoassociated
molecules in the $^{1}\Pi_{g}$ state for a range of laser intensities,
cf. Fig.~\ref{fig:intensity_j}. For weak fields, the purity
is roughly constant as a function of intensity and about three times
larger than the purity obtained for the intensity of $5\times
10^{12}\,$W/cm$^2$ used in the experiment. As intensity is increased,
a drop in the purity is observed which levels off at large
intensities. We attribute this drop to power broadening for strong
fields, which brings more atom pairs into the Franck-Condon window for
PA.

The purity of the photoassociated sub-ensemble is less than the
inverse of the number of occupied energy states
due to coherence. To quantify this effect  we separate static and dynamic
contributions, $\Op\rho = \Op\rho_{stat} +\Op\rho_{dyn}$.
Expressing the density operator $\Op\rho$ in the energy representation
the static part corresponds to the diagonal matrix elements and the
dynamical coherence to the off-diagonal elements.
The dynamical contributions are quantified by the
coherence measure ${\cal C} =
\Tr[\Op\rho^2_{dyn}(t)]$.\cite{BaninJCP94}. 
Figure~\ref{fig:intensity_j} shows
the coherence measure of the excited state, ${\cal C}^e$,
as a function of laser intensity (red diamonds). It is found to be
about one order of magnitude smaller than the purity.
This is rationalized by the change in Franck-Condon points with
different $J$ which degrade the vibrational coherence.
Within a sub-ensemble for a given angular momentum $J$
the difference between ${\cal C}_J^e$ and ${\cal P}_J^e$
is less than an order of magnitude.

\section{Conclusions}
\label{sec:concl}

We have described two-photon femtosecond photoassociation of magnesium
atoms from first principles using state of the art \textit{ab initio}
methods and quantum dynamical calculations.
Highly accurate potential energy curves were obtained using the
coupled cluster method and a large basis set. Two-photon couplings and
dynamic Stark shifts are important to correctly model the
interaction of the atom pairs with the strong field of a
femtosecond laser pulse. They
were calculated within the framework of the equation of motion
(response) coupled cluster method.
The photoassociation dynamics were obtained by solving the
time-dependent Schr\"odinger equation for all relevant partial waves,
accounting for the laser-matter interaction in a non-perturbative way,
and performing a thermal average.

We have developed an efficient numerical method to describe the
incoherent thermal ensemble that is the initial state for
photoassociation at high temperatures. It is based on random phase
wave packets which can be built from eigenfunctions of the grid,
the Hamiltonian, or the kinetic energy. The latter can provide a rough
estimate which is sufficient
to yield qualitatively correct results. It
neglects, however, the contribution from
bound levels and long-lived shape resonances and therefore comes with
error bars of about 20\%. The best compromise
between high accuracy and convergence is found for the
eigenfunction-based method where random phase realizations are built
from the eigenfunctions of the electronic ground state
Hamiltonian. About 200 partial waves and 200 realizations for each
partial wave are required for converged photoassociation dynamics.
Time-dependent thermal averages are
obtained by propagating each of the random phase wave functions and
incoherently summing up all single expectation values.

The random phase approach allows for constructing the thermal atom pair
density as a function of interatomic separation for high
temperatures. This is important to highlight the difference between
hot and cold
photoassociation.\cite{VardiJCP97,JiriPRA01,MyJPhysB06} In the cold
regime, the largest density is defined by the quantum reflection and
resides  in the long-distance, downhill part of the potential. The
opposite is true in the hot regime: Here, the largest density is found
in the repulsive part of the ground state potential. This is due to
the many partial waves that are
thermally populated and the colliding atom pairs having sufficiently
high kinetic energy to overcome the rotational barriers.
For specific partial waves, shape resonances are found to play a role.
This is not surprising since they represent quasi-bound
states that are ideally suited for
photoassociation.\cite{GonzalezKochPRA12}
At very low temperatures, most partial waves are frozen out and the
scattering is  almost exclusively $s$-wave.
The role of the rotational quantum
numbers $J$ is less important in the electronically excited state but
it is still detectable in form of quantum beats.\cite{RybakPRL11}
Both hot and cold photoassociation come with advantages as well as
drawbacks. In the hot regime, molecules with much shorter bond length
than in the cold regime are formed.  However, the quantum purity and
coherence of the created molecules is much larger in the cold regime
where dynamical correlations exist prior to photoassociation. These
correlations indicate pre-entanglement of the atom pair. Making a
molecule corresponds to entangling two atoms, and photoassociation
amounts to filtering out an entangled subensemble both in the hot and
cold regime.

Our work has opened up the possibility to study femtosecond
photoassociation and its control at high temperatures and
to investigate systematically the generation of coherence out of an
incoherent initial state. Future efforts will address the efficient
theoretical description of the probe step. The theme of coherent
control of binary reactions requires a sound theoretical basis to
which our current study lays the ground work.

\section*{Acknowledgments}
This study was supported by the Israeli Science Foundation ISF Grant
No. 1450/10, by the Deutsche Forschungsgemeinschaft and in part
by the National Science Foundation under Grant No. NSF PHY11-25915.
CPK, DMR, MT, RK and RM enjoyed hospitality of KITP.
RM and MT would like to thank the Polish Ministry of Science and
Higher Education for the financial support through the project
N N204 215539. MT was supported by the project operated within
the Foundation for Polish Science MPD Programme co-financed by
the EU European Regional Development Fund.

\appendix

\section{Classical approximation of the partition function}
\label{sec:app}

The classical approximation of the partition function  is obtained
starting from the standard definition,
\[
  Z_{cl} = \frac{1}{h^3} \int d^3 R \int d^3 P
  e^{-\beta\left(\frac{\vec{P}^2}{2m}+V(R)\right)}\,.
\]
Performing the integral over angles and introducing polar momentum
coordinates, we find
\begin{eqnarray*}
  Z_{cl} &=&\frac{4\pi}{h^3} \int_0^{R_{max}} dR R^2 e^{-\beta V(R)}\\
  &&\quad\quad 2\pi \int_{-\infty}^\infty dP_R\int_{-\infty}^\infty
  dP_\perp P_{\perp} e^{-\frac{\beta}{2m}(P_R^2+P_{\perp}^2)}\\
  &=& \frac{4\pi^2}{h^3} \int_0^\infty 2J dJ
  \int_0^{R_{max}} dR e^{-\beta \left(V(R)+\frac{J^2}{2mR^2}\right)}\\
  &&\quad\quad\int_{-\infty}^\infty dP_R
  e^{-\beta\frac{P_R^2}{2m}}\,,
\end{eqnarray*}
where we made use of $J=R P_\perp$. Carrying out the integral over
the radial momentum yields
\[
  Z_{cl} = \frac{4\pi^2}{h^3}\sqrt{\frac{2m\pi}{\beta}} \int 2J dJ
  \int dR e^{-\beta \left(V(R)+\frac{J^2}{2mR^2}\right)}\,.
\]
Approximating the potential $V(R)\approx 0$, the integral over the
computational box of size $R_{max}$ can be performed,
\begin{eqnarray*}
  Z_{cl} &=& \frac{4\pi^2}{h^3}\sqrt{\frac{2m\pi}{\beta}} \int dJ\,2J\, Z_J^{R_{max}}
\end{eqnarray*}
with
\begin{eqnarray*}
Z_J^{R_{max}} &=& \sqrt{\frac{\pi\beta J^2}{2m}} \left[
  \mathrm{erf}\left( \frac{1}{R_{max}}\sqrt{\frac{\pi\beta J^2}{2m}}\right)-1
\right] \\ &&+ R_{max} e^{-\beta\frac{J^2}{2mR_{max}^2}}\,.
\end{eqnarray*}


\end{document}